\documentstyle[11pt,amssymb]{article}

  \let\n=\nu

\let\C=\Chi

\def\nn{\nonumber} \def\bd{\begin{document}} \def\ed{\end{document}}
\def\ds{\documentstyle} \let\fr=\frac \let\bl=\bigl \let\br=\bigr
\let\Br=\Bigr \let\Bl=\Bigl
\let\bm=\bibitem
\let\na=\nabla
\let\pa=\partial \let\ov=\overline
\newcommand{\be}{\begin{equation}}
\newcommand{\ee}{\end{equation}}
\def\ba{\begin{array}}
\def\ea{\end{array}}
\def\ft#1#2{{\textstyle{{\scriptstyle #1}\over {\scriptstyle #2}}}}
\def\fft#1#2{{#1 \over #2}}
\def\del{\partial}
\def\vp{\varphi}
\def\st#1{{\scriptstyle #1}}
\def\sst#1{{\scriptscriptstyle #1}}

\def\oneone{\rlap 1\mkern4mu{\rm l}}
\def\td{\tilde}
\def\wtd{\widetilde}
\def\ie{\rm i.e.\ }
\def\dalemb#1#2{{\vbox{\hrule height .#2pt
        \hbox{\vrule width.#2pt height#1pt \kern#1pt
                \vrule width.#2pt}
        \hrule height.#2pt}}}
\def\square{\mathord{\dalemb{6.8}{7}\hbox{\hskip1pt}}}

\def\cramp{\medmuskip = 2mu plus 1mu minus 2mu}
\def\cramper{\medmuskip = 2mu plus 1mu minus 2mu}
\def\crampest{\medmuskip = 1mu plus 1mu minus 1mu}
\def\uncramp{\medmuskip = 4mu plus 2mu minus 4mu}

\newcommand{\ho}[1]{$\, ^{#1}$}
\newcommand{\hoch}[1]{$\, ^{#1}$}
\newcommand{\bea}{\begin{eqnarray}}
\newcommand{\eea}{\end{eqnarray}}
\newcommand{\ra}{\rightarrow}
\newcommand{\lra}{\longrightarrow}
\newcommand{\Lra}{\Leftrightarrow}
\newcommand{\ap}{\alpha^\prime}
\newcommand{\bp}{\tilde \beta^\prime}
\newcommand{\tr}{{\rm tr} }
\newcommand{\Tr}{{\rm Tr} }
\def\0{{\sst{(0)}}}
\def\1{{\sst{(1)}}}
\def\2{{\sst{(2)}}}
\def\3{{\sst{(3)}}}
\def\4{{\sst{(4)}}}
\def\5{{\sst{(5)}}}
\def\6{{\sst{(6)}}}
\def\7{{\sst{(7)}}}
\def\8{{\sst{(8)}}}
\def\n{{\sst{(n)}}}
\def\cA{{{\cal A}}}
\def\cF{{{\cal F}}}
\def\tV{\widetilde V}
\def\tW{\widetilde W}
\def\tH{\widetilde H}
\def\tE{\widetilde E}
\def\tF{\widetilde F}
\def\tA{\widetilde A}
\def\im{{{\rm i}}}
\def\jm{{{\rm j}}}
\def\km{{{\rm k}}}

\def\tY{{{\wtd Y}}}
\def\ep{{\epsilon}}
\def\vep{{\varepsilon}}
\def\R{\rlap{\rm I}\mkern3mu{\rm R}}
\def\bD{{{\bar D}}}
\def\R{{{\Bbb R}}}
\def\C{{{\Bbb C}}}
\def\H{{{\Bbb H}}}
\def\CP{{{\Bbb C}{\Bbb P}}}
\def\RP{{{\Bbb R}{\Bbb P}}}
\def\Z{{{\Bbb Z}}}
\def\bA{{{\Bbb A}}}
\def\bB{{{\Bbb B}}}

\newcommand{\NP}{Nucl. Phys. }
\newcommand{\tamphys}{\it Center for Theoretical Physics,
Texas A\&M University, College Station, TX 77843, USA}
\newcommand{\umich}{\it Michigan Center for Theoretical Physics,
University of Michigan\\ Ann Arbor, MI 48109, USA}
\newcommand{\upenn}{\it Department of Physics and Astronomy,
University of Pennsylvania\\ Philadelphia,  PA 19104, USA}
\newcommand{\SISSA}{\it  SISSA-ISAS and INFN, Sezione di Trieste\\
Via Beirut 2-4, I-34013, Trieste, Italy}

\newcommand{\ihp}{\it Institut Henri Poincar\'e\\
  11 rue Pierre et Marie Curie, F 75231 Paris Cedex 05}

\newcommand{\damtp}{\it DAMTP, Centre for Mathematical Sciences,
 Cambridge University\\ Wilberforce Road, Cambridge CB3 OWA, UK}
\newcommand{\itp}{\it Institute for Theoretical Physics, University of
California\\ Santa Barbara, CA 93106, USA}

\newcommand{\auth}{M. Cveti\v{c}\hoch{\dagger}, G.W. Gibbons\hoch{\sharp},
James T. Liu\hoch{\star}, H. L\"u\hoch{\star} and C.N. Pope\hoch{\ddagger}}

\begin{document}
\begin{flushright}
\hfill{DAMTP-2001-56}\ \ \ {CTP TAMU-22/01}\ \ \ {UPR-946-T}\ \ \
{MCTP-01-28}\\
{June 2001}\ \ \
{hep-th/0106162}
\end{flushright}


\begin{center}
{ \large {\bf A New Fractional D2-brane, $G_2$ Holonomy and T-duality}}

\vspace{5pt}
\auth

\vspace{3pt}
{\hoch{\dagger}\upenn}

\vspace{3pt}
{\hoch{\sharp}\damtp}

\vspace{3pt}
{\hoch{\star}\umich}

\vspace{3pt}
{\hoch{\ddagger}\tamphys}

\vspace{3pt}

\underline{ABSTRACT}
\end{center}

    Recently, a new example of a complete non-compact Ricci-flat
metric of $G_2$ holonomy was constructed, which has an asymptotically
locally conical structure at infinity with a circular direction whose
radius stabilises.  In this paper we find a regular harmonic 3-form in
this metric, which we then use in order to obtain an explicit solution
for a fractional D2-brane configuration.  By performing a T-duality
transformation on the stabilised circle, we obtain the type IIB
description of the fractional brane, which now corresponds to D3-brane
with one of its world-volume directions wrapped around the circle.


\pagebreak
\setcounter{page}{1}

\vfill\eject

\section{Introduction}

    The T-duality that relates the type IIA and type IIB strings
allows one to establish a mapping between a D$p$-brane in one theory
and a D$(p+1)$-brane in the other.  Recently, more general classes of
D$p$-brane solution have been constructed, in which further fluxes for
form fields are present, in addition to the one that carries the
standard D$p$-brane charge.  An example of this type is the fractional
D3-brane of Klebanov and Strassler \cite{klebstra}.  Subsequently, a
number of other fractional or resolved brane solutions have been
constructed, including fractional D2-branes in which the usual flat
transverse 7-metric is replaced by a smooth Ricci-flat metric of $G_2$
holonomy \cite{clptrans,d2nsns2}.  It therefore becomes of interest to
study the possibility of relating fractional branes, such as these
D2-branes of type IIA, to branes in the T-dual type IIB description.

   New families of complete eight-dimensional manifolds of Spin(7)
holonomy have recently been constructed, which are asymptotically
locally conical (ALC) \cite{newspin7}.  In these, the geometry at
large distance locally approaches the product of a circle of fixed
radius, and an asymptotically conical (AC) 7-manifold.  Deformed
M2-branes using these Spin(7) manifolds were constructed in
\cite{newspin7}, and it was shown that dimensional reduction on the
circle gives D2-brane solutions in $D=10$ that are similar to those
that can be built using $G_2$ manifolds.

   This Spin(7) construction motivated an analogous construction of
asymptotically locally conical seven-dimensional manifolds of $G_2$
holonomy, and an explicit isolated example has been obtained
\cite{bggg}.  It has the same property as the Spin(7) example, of
locally approaching the product of a circle of fixed radius, and an
asymptotically conical manifold (of dimension 6 in this case), at
large distance.

    In this paper, we shall construct a new fractional D2-brane
solution, using the new ALC metric of $G_2$ holonomy obtained in
\cite{bggg}.  In order to obtain the fractional D2-brane using this
metric, we first need to construct a suitable well-behaved harmonic
3-form $G_\3$, which is regular at short distance and which falls off
at infinity.  The integral of $|G_\3|^2$ diverges logarithmically
with proper distance.  (This is reminiscent of the behaviour of the
harmonic 3-form used in the construction of the fractional D3-brane
\cite{klebstra}.)

    Having obtained the harmonic 3-form, we then use it in order to
construct the associated fractional D2-brane solution, which is
supersymmetric.  Remarkably, it turns out that the equations can be
exactly integrated, and so we are able to give the result explicitly
in closed form.

   The fact that the new $G_2$ metric of \cite{bggg} is asymptotically
locally conical means that the circle whose radius tends to a constant
at infinity is a natural candidate on which to perform a T-duality
transformation.  We carry out this transformation, and show how the
fractional D2-brane is mapped into a supersymmetric wrapped D3-brane
in the type IIB theory.  In order to set the scene for this duality
mapping, we first carry it out for the simpler situation of a
``vacuum'' (Minkowski)$_3\times {\cal M}_7$, where ${\cal M}_7$
denotes the Ricci-flat $G_2$ manifold.

       The paper is organised as the following.  In section 2, we
review the non-compact cohomogeneity one $G_2$ manifolds and
previously known fractional D2-branes.  In section 3, we review the
recently constructed new $G_2$ manifolds.  In section 4, we obtain the
regular harmonic 3-form for the new $G_2$ manifold and construct
new fractional D2-branes.  We perform T-duality and obtain the $S^1$
wrapped fractional D3-brane in section 5.  Finally, we conclude the
paper in section 6.

\section{Review of $G_2$ manifolds and fractional D2-branes}

     Three explicit metrics of cohomogeneity one for seven-dimensional
manifolds with $G_2$ holonomy have been known for some time
\cite{brysal,gibpagpop}.  The first two have principal orbits that are
$\CP^3$ or $SU(3)/(U(1)\times U(1))$, viewed as an $S^2$ bundle over
$S^4$ or $\CP^2$ respectively.  The associated 7-manifolds have the
topology of an $R^\3$ bundle over $S^4$ or $\CP^2$.  The third
manifold has principal orbits that are topologically $S^3\times S^3$,
viewed as an $S^3$ bundle over $S^3$, and the 7-manifold is
topologically $\R^4\times S^3$.  Recently, there has been a
considerable interest in studying $D=4$, ${\cal N}=1$ theory from the
compactification of M-theory on $G_2$ holonomy spaces
\cite{acharya,amv,wit-talk,aw,gomis,en,kacmcg,gutpap,%
kkp,av,bggg,hern,dasohtat}.  This provides a geometric description of
M-theory on $G_2$ manifolds for the strongly coupled $D=4$, ${\cal
N}=1$ dual field theory arising from the wrapped D6-brane on
conifolds.

        Supersymmetric M3-branes, arising as deformations of
(Minkowski)$_4$ times ($G_2$ holonomy) backgrounds, with non-vanishing
4-form field in eleven-dimensional supergravity, were constructed in
\cite{m3brane1,m3brane2}.

      Another natural application for the spaces of $G_2$ holonomy is
in the construction of fractional D2-branes, since they have
non-vanishing Betti numbers $b_3$ and $b_4$.  The fractional D2-brane
solutions of type IIA supergravity using the three above-mentioned
spaces of $G_2$ holonomy are given by \cite{clptrans,d2nsns2}
\bea
ds_{10}^2 &=& H^{-5/8}\, dx^\mu\, dx^\nu\, \eta_{\mu\nu} +
    H^{3/8}\, ds_7^2\,,\nn\\
F_\4&=& d^3x\wedge dH^{-1} + m\, G_\4\,,\qquad
F_\3=m\, G_\3\,,\qquad \phi=\ft14\, \log H\,,\label{d2sol}
\eea
where $G_\3$ is an harmonic 3-form in the Ricci-flat 7-metric
$ds_7^2$, and $G_\4=*G_\3$, with $*$ the Hodge dual with respect to
$ds_7^2$.  The function $H$ satisfies
\be
\square H=-\ft16 m^2\, G_\3^2\,,\label{Heq}
\ee
where $\square$ denotes the scalar Laplacian in the
transverse 7-metric $ds_7^2$.

        The fractional D2-brane for the $\R^4$ bundle over $S^3$ was
constructed in \cite{clptrans}, and its supersymmetry was demonstrated
in \cite{d2nsns2}.  The harmonic 3-form is square integrable at small
distance, but linearly non-normalisable at large distance.
Correspondingly, the solution is regular everywhere, with the small
distance structure (Minkowski)$_3\times \R^4\times S^3$, whilst at
large distance, the function $H$ behaves like \cite{clptrans}
\be
H\sim 1 + \fft{m^2}{4r^4} -\fft{4m^2}{15r^5}\,.
\ee
Note that this leading-order $1/r^4$ behaviour is more like that for a
standard D3-brane, which has a six-dimensional transverse space,
rather than usual $1/r^5$ behaviour for the seven-dimensional
transverse space that we have here.

        The fractional D2-branes in which the transverse space is the
$R^3$ bundle over $S^4$ or $\CP^2$ were constructed in \cite{d2nsns2}.
In these cases, the harmonic 4-form is $L^2$-normalisable, and
consequently, the solution is regular everywhere.  Its large-distance
asymptotic behaviour is
\be
H\sim  1+ \fft{c\, m^2}{r^5} -\fft{m^2}{4r^6}\,,
\ee
where $c$ is a certain numerical constant.

\section{Review of the new $G_2$ manifolds}

       Recently, more general metrics with the structure of an $R^4$
bundle over $S^3$ have been considered \cite{bggg,m3brane2}, with a
view to obtaining further examples of metrics of $G_2$ holonomy.  This
would be analogous to the recent construction of more general
eight-dimensional metrics of Spin(7) holonomy in \cite{newspin7}.  One
can consider the following ansatz for seven-dimensional metrics:
\be
ds_7^2 = h^2\, dr^2 + a_i^2\, \td h_i^2 + b_i^2\, h_i^2\,,
\label{sixpara}
\ee
where
\be
h_i \equiv \sigma_i + \Sigma_i\,,\qquad \td h_i = \sigma_i
-\Sigma_i\,.
\ee
Here $\sigma_i$ and $\Sigma_i$ are the left-invariant 1-forms on two
$SU(2)$ group manifolds, $S^3_\sigma$ and $S^3_\Sigma$, and $a_i$,
$b_i$ and $h$ are functions of the radial coordinate $r$.  The
principal orbits are therefore $S^3$ bundles over $S^3$, and since the
bundle is topologically trivial, they have the topology $S^3\times
S^3$.  The ansatz is a generalisation of the one in
\cite{brysal,gibpagpop} that was used for obtaining the original
complete metric of $G_2$ holonomy on the $\R^4$ bundle over $S^3$.

     It was shown in \cite{m3brane2,bggg} that the conditions for
Ricci flatness for the ansatz (\ref{sixpara}) admit as first integrals
a system of first-order equations that can be derived from a
superpotential.  In \cite{m3brane2} it was shown that these
first-order equations are the integrability conditions for the
existence of a covariantly-constant spinor, and hence for $G_2$
holonomy.  An equivalent demonstration of $G_2$ holonomy was given in
\cite{bggg}, by showing that the first-order equations follow by
requiring the covariant constancy of a 3-form.  The general explicit
solution is not known.  If one specialises to $a_i=a$ and $b_i=b$,
then the general solution becomes the previously known $G_2$ manifold.

   In \cite{bggg} a less restrictive specialisation was made,
involving setting $a_1=a_2$ and $b_1=b_2$.  Writing the ansatz now as
\be
ds_7^2 = h^2\, dr^2 + a^2\, (\td h_1^2 + \td h_2^2) + c^2\, \td h_3^2
 + b^2\, (h_1^2 + h_2^2) + f^2 \, h_3^2\,,\label{fourpara}
\ee
one again obtains first-order equations, which have not been solved in
general.  However, a special solution was found in \cite{bggg}, which
may be written as
\bea a&=& \fft1{\sqrt{8\ell}}\, \sqrt{(r-\ell)(r+3\ell)}\,,\qquad
b= \fft1{\sqrt{8\ell}}\, \sqrt{(r+\ell)(r-3\ell)}\,\nn\\
c &=& -\fft{r}{\sqrt{6\ell}}\,,\qquad
f = \sqrt{\fft{2\ell}{3}}\,
   \fft{\sqrt{r^2-9\ell^2}}{\sqrt{r^2-\ell^2}}\,,\label{simsol}
\eea
with $h=1/f$.  (Note that the minus sign in the expression for $c$ is
purely conventional.) The radial coordinate runs from $r=3\ell$ to
infinity.  The metric (\ref{fourpara}) is then complete on a manifold
with the same $\R^4\times S^3$ topology as the original metric in
\cite{brysal,gibpagpop}.  However, an important difference is that in
the new metric the radius in the $S^1$ direction associated with $h_3$
becomes a constant asymptotically at large $r$.  Thus the metric is no
longer asymptotically conical, but instead it locally approaches the
product of a circle and an asymptotically-conical six-metric at
infinity.  This is analogous to the new Spin(7) manifolds obtained in
\cite{newspin7}.  Arguments were also given in \cite{bggg} for the
existence of more general solutions of the first-order equations
for the $G_2$ metrics (7), which would be analogous to the more
general Spin(7) solutions in \cite{newspin7}.

\section{New fractional D2-brane}

\subsection{Harmonic 3-form}

   As was discussed in section 2, the construction of a fractional
D2-brane requires a well-behaved harmonic 3-form in the $G_2$ holonomy
space. In this subsection, we construct a harmonic 3-form $G_\3$ in
the new $G_2$ metric. We begin by choosing the natural vielbein
\be
e^0=h\, dr\,,\quad e^1=a\, \td h_1\,,\quad e^2=a\,
\td h_2\,,\quad e^3 = c\,
\td h_3\,,\quad e^4 = b\, h_1\,,\quad e^5=b\, h_2\,,\quad
e^6 = f\,h_3
\ee
for the metric (\ref{fourpara}).
Motivated by the ansatz for the harmonic form in the original $G_2$
metric on $\R^4\times S^3$ \cite{clptrans,d2nsns2,m3brane2}, we make
the following ansatz for the 4-form $G_\4={*G_\3}$ dual to $G_\3$:
\bea
G_\4 &=& u_1\, e^1\wedge e^2\wedge e^4\wedge e^5 + u_2\,
     e^2\wedge e^3\wedge e^5\wedge e^6 + u_2\,
     e^3\wedge e^1 \wedge e^6\wedge e^4 \nn\\
&&+ u_3\, e^0\wedge e^4\wedge e^5\wedge e^6 + u_4\, e^0\wedge
e^1\wedge e^2\wedge e^6 \nn\\
&&+ u_5\, e^0\wedge e^2\wedge e^3\wedge e^4
  + u_5\, e^0\wedge e^3\wedge e^1\wedge e^5\,.\label{4fans}
\eea
(The ansatz for the original $G_2$ metric is like this, with
$u_2=u_1$, and $u_5=u_4$.)   Its Hodge dual is given by
\bea
G_\3 &=& -u_1\, e^0\wedge e^3\wedge e^6 - u_2\, e^0\wedge e^1\wedge
e^4 - u_2\, e^0\wedge e^2\wedge e^5 + u_3\,
e^1\wedge e^2\wedge e^3\nn\\
&&+ u_4\, e^3\wedge e^4\wedge e^5 +
u_5\, e^1\wedge e^5\wedge e^6 - u_5\, e^2\wedge e^4\wedge e^6\,.
\label{f3exp}
\eea

   After straightforward manipulations, we can obtain the first-order
equations that follow from imposing $dG_\4=0$ and $d{*G_\4}=0$.  In
order to simplify the task of solving these, it is useful to note that
in previous examples, the harmonic forms that could give rise to
supersymmetric fractional branes all had the feature that some
constant linear combination of the functions $u_i$ in the ansatz for
the harmonic form vanished
\cite{klebstra,clptrans,cglpsten,d2nsns2,cglphyper}.  We find that in
this case too, the system of first-order equations following from
$dG_\4=0$ and $d{*G_\4}=0$ imply that a certain linear combination of
the $u_i$ functions in (\ref{4fans}) is a constant, which we can
choose to be zero.  Thus we are led to impose
\be
u_1 -2 u_2 -u_3 + u_4 -2 u_5=0\,.
\ee
This linear relation ensures that the fractional D2-brane is
supersymmetric. We are left with four first-order equations for the
remaining undetermined functions, and so the general solution has four
constants of integration.  Requiring that the 4-form be well-behaved
at the origin $r=3\ell$, and that it fall off at large $r$, fixes
these integration constants completely, up to an overall scale.  We
then find that the functions $u_i$ are given by
\bea
u_1 &=& \fft{4[1053 - 441 r^2  + 27 r^4  + r^6  -
                36 (-9 - r^2  + 2 r^4 ) \, \log(\ft{r}{3})]}{
           (r^2-9)^3\, (r^2-1)^2}\,,\nn\\
u_2 &=& \fft{3(r^2+3)\, [r^4-81 -36 r^2\, \log(\ft{r}{3})]}{
                         r^2\, (r^2-9)^3\, (r^2-1)}\,,\nn\\
u_3 &=& \fft{-(r^2-9)(-9+15r+17r^2+15r^3+2r^4) +
    36 r^2\, (3r+1)\, \log(\ft{r}{3})}{(r+3)^3\, (r+1)\,r^2\, (r-1)^2 (r-3)}
\,,\nn\\
u_4 &=&  \fft{(r^2-9)(9+15r-17r^2+15r^3-2r^4) - 36r^2\, (3r-1)\,
      \log(\ft{r}{3})}{(r-3)^3\, (r-1)\,r^2\, (r+1)^2 (r+3)}\,,\nn\\
u_5 &=& \fft{4[ r^4 -81  - 36 r^2\, \log(\ft{r}{3})]}{(r^2-9)^3\, (r^2-1)}\,.
\label{usol}
\eea
Note that without losing generality, we have set the scale parameter
$\ell$ in the metric functions in (\ref{simsol}) to $\ell=1$ here, to
simplify the writing somewhat.  It should be remarked that despite
appearances, the functions $u_i$ in (\ref{usol}) are actually
non-singular at the minimum radius $r=3$, and they have regular Taylor
expansions there.  Note that the harmonic 3-form we have constructed
here, which is localised near the origin, is quite distinct from the
covariantly-constant calibrating 3-form discussed in \cite{bggg}.

  The magnitude of $G_\3$ is given by
\be
|G_\3|^2 = 6 (u_1^2 + 2 u_2^2 + u_3^2 + u_4^2 + 2 u_5^2)\,.
\ee
Substituting the above expressions for the $u_i$ into this, we find
that at short distance, near to $r=3$, we have
\be
|G_\3|^2 = \fft{7}{81} - \fft{49(r-3)}{1243} + \fft{4973
(r-3)^2}{17496} + \cdots\,.
\ee
At large distance, we have
\be
|G_\3|^2 = \fft{48}{r^6} + \fft{120}{r^8} + \fft{96[241 - 252
\log(\ft{r}{3})]}{r^{10}} + \cdots\,.
\ee
It follows from this, and the fact that $\sqrt{g}\sim r^5$ at large
$r$, that $\int \sqrt{g}\, |G_\4|^2$ diverges logarithmically at large
$r$.  The harmonic 4-form is therefore not $L^2$ normalisable, and the
rather slow logarithmic divergence in the integral of $|G_\4|^2$ is
very similar to that encountered in the fractional D3-brane
construction of Klebanov and Strassler \cite{klebstra}.

\subsection{New fractional D2-brane}

   In this section we show that the metric function $H$ in the
deformed D2-brane solution (\ref{d2sol}),(\ref{Heq}) using this
harmonic form can be obtained explicitly.  First we note that
\bea
\int_3^r \sqrt{g}|G_\3|^2 \!\!\!&=&\!\!\!
\fft{\sqrt6\, (2r^{8} - 49r^6 + 729 r^4 +1161
r^2 -243)}{32\, r^2\, (r^2-3)\,(r^2+3)^2\, (r^2-1)^2}\nn\\
\!\!\!&&\!\!\!-\fft{\sqrt6}{8(r^2-9)^4\, (r^2-1)^2}\Big(-324r^2\, (11r^4
+2r^2-45)\, (\log(\ft{r}{3}))^2\\
\!\!\!&&\!\!\!+(r^2-9)\, (r^{10} -29 r^8 + 424r^6 + 1944 r^4 -5913 r^2
-2187)\, \log(\ft{r}{3})\Big)\,.\nn
\eea
From this, we find that $H$ is given by
\bea
H &=& c_0 + \fft{m^2\, (22r^8 -589 r^6 + 5247 r^4 -19035 r^2
-3645)}{30 r^2\, (r^2-9)^3\, (r^2-1)} \nn\\
&&
+ \fft{2m^2\, (11 r^8 -335 r^6 +3645 r^4 -14661 r^2 +43740)}{15
(r^2-9)^4\, (r^2-1)}\, \log(\ft{r}{3})\nn\\
&&  -\fft{1944 m^2\, (11r^2+9)}{5(r^2-9)^5\, (r^2-1)}\,
(\log(\ft{r}{3}))^2
\nn\\
&&+
\fft{22 m^2}{135}\, [\psi(-\ft{r}{3}) - \psi(1-\ft{r}{3}) +
\log(\ft{r}{3})\, \log(1+\ft{r}{3}) -
(\log(\ft{r}{3}))^2 ]\,,
\eea
where $c_0$ is a constant and
$\psi(x)\equiv -\int_{0}^x y^{-1}\, \log(1-y)\, dy$ is the
dilogarithm.

   The function $H$ becomes a constant at small distance $r\rightarrow
3$, and at large distance it has the asymptotic form
\be
H=c_0 + \fft{3m^2\,(4\log(\ft{r}{3}) -1)}{4r^4} +
\fft{3m^2\,(24\log(\ft{r}{3})-3)}{2r^6} +
\fft{m^2\,(8856\log(\ft{r}{3})-2229)}{16r^8} +\cdots\,.\label{hlarge}
\ee
The solution is supersymmetric, and regular everywhere, with no
horizon.  It does not give a well-defined ADM mass per unit 3-volume;
we have $M\sim r^5\, H'$ in the limit $r\longrightarrow\infty$, and
this diverges logarithmically.  The situation is directly analogous to
that in the fractional D3-brane \cite{klebstra}.

   The NS-NS 3-form carries a non-vanishing magnetic charge.  The
integral of $F_\3$ over the non-trivial 3-cycle can conveniently be
evaluated by restricting it to the $S^3$ bolt at $r=3\ell$ (which
means $r=3$ since we have set $\ell=1$).  From (\ref{f3exp}) we have
\be
F_\3\Big|_{r=3} = m\, G_\3\Big|_{r=3} =
\fft{m}{2\sqrt6}\, \td h_1\wedge \td h_2\wedge \td h_3\,,
\ee
and hence we find $\int_{S^3}\, F_\3 = 64 m\, \pi^2/\sqrt6$.

   The ``standard'' 4-form electric flux is given by integrating the
quantity $d(e^{\fft12\phi}\, {*F_\4})$ over a 7-volume bounded by a
6-surface ${\cal M}_6$.  Since $d(e^{\fft12\phi}\, {*F_\4}) =
F_\3\wedge F_\4$, the electric flux can be written as $\int F_\3\wedge
F_\4$.  From (\ref{d2sol}), this gives $m^2\, \int G_\3\wedge G_\4$,
which is nothing but $12 m^2\, \int |G_\3|^2$.  As noted above, $G_\3$
is not $L^2$ normalisable, and so if the 7-volume is taken to be the
interior of the level set $r=$~constant, the integral diverges
logarithmically with proper distance.  This is precisely what one
expects, since the solution is supersymmetric and, as we saw above,
the ADM mass per unit 3-volume diverges logarithmically too.

\section{T-duality}

     The new $G_2$ metric (\ref{fourpara}) has a circle whose radius
is stabilised at infinity.  To see this explicitly, let us first write
the $SU(2)$ left-invariant 1-forms $\sigma_i$ and $\Sigma_i$ in terms
of Euler angles
\bea
&&\sigma_1 = \cos\psi_1\, d\theta_1 +
\sin\psi_1\,\sin\theta_1\, d\phi_1\,,\qquad
\Sigma_1 = \cos\psi_2\, d\theta_2 +
\sin\psi_2\,\sin\theta_2\, d\phi_2\,,\nn\\
&&
\sigma_2 = -\sin\psi_1\, d\theta_1 +
\cos\psi_1\,\sin\theta_1\, d\phi_1\,,\qquad
\Sigma_2 = -\sin\psi_2\, d\theta_2 +
\cos\psi_2\,\sin\theta_2\, d\phi_2\,,\nn\\
&&
\sigma_3 = d\psi_1 + \cos\theta_1\, d\phi_1\,,\qquad
\Sigma_3 = d\psi_2 + \cos\theta_2\, d\phi_2\,.
\eea
Then the combination of $(\psi_1 + \psi_2)$ appears in the metric
(\ref{fourpara}) only in $h_3$, as $d(\psi_1+\psi_2)$.
Specifically, after making the following redefinition,
\be
\psi=\psi_1 + \psi_2\,,\qquad
\td \psi = \psi_1-\psi_2\,,
\ee
we have
\bea
ds_7^2 &\equiv& ds_6^2 + f^2\, h_3^2\nn\\
&=& h^2\, dr^2 + (a^2 + b^2)\, \Big(d\theta_1^2 +d\theta_2^2 +
\sin^2\theta_1\, d\phi_1^2 + \sin^2\theta_2\, d\phi_2^2\Big)\nn\\
&&+2(a^2 -b^2)\Big(\sin\td\psi\, (\sin\theta_1\, d\theta_2\,d\phi_1 -
\sin\theta_2\, d\theta_1\, d\phi_2) \nn\\
&&\qquad\qquad\qquad+
\cos\td\psi\, (d\theta_1\, d\theta_2 + \sin\theta_1\, \sin\theta_2\,
d\phi_1\, d\phi_2)\Big)\nn\\
&& +c^2\, (d\td\psi + \cos\theta_1\, d\phi_1 -\cos\theta_2\, d\phi_2)^2
+ f^2 (d\psi + \cos\theta_1\, d\phi_1 +\cos\theta_2\, d\phi_2)^2\,.
\label{eulermetric}
\eea
From this, it can be seen that there are three commuting $U(1)$
Killing vector fields of the metric, namely $\del/\del\phi_1$,
$\del/\del\phi_2$ and $\del/\del\psi$.  The Killing vectors
$\del/\del\phi_1$ and $\del/\del\phi_2$ generate $U(1)$ subgroups
$U(1)_L^\sigma$ and $U(1)_L^\Sigma$ of the left-acting
$SU(2)_L^\sigma$ and $SU(2)_L^\Sigma$ groups in the two 3-spheres.
The Killing vector $\del/\del\psi$ generates the subgroup
$U(1)_D$ of the diagonal right-acting $SU(2)_R^D$.

   If the $G_2$ manifold ${\cal M}_7$ is taken as a factor in an
M-theory vacuum (Minkowski)$_4\times {\cal M}_7$, then we can reduce on
any of the three circles $U(1)_L^\sigma$, $U(1)_L^\Sigma$ or $U(1)_D$
to give D6-branes in type IIA theory.  In particular, the circle
$U(1)_D$ with stablised radius gives a distance-independent R-R vector
potential
\be
{\cal A}_\1 = \cos\theta_1\, d\phi_1 +\cos\theta_2\, d\phi_2\,,
\ee
implying that
\be
{\cal F}_\2 =d{\cal A}_1 = J_\2 \equiv\Omega^1_\2 +\Omega^2_\2\,.
\ee
(This reduction of the metric (\ref{fourpara}) with the solution
(\ref{simsol}) was discussed in \cite{bggg}.)

    If the $G_2$ manifold ${\cal M}_7$ is instead taken as a factor in
a $D=10$ vacuum (Minkowski)$_3\times {\cal M}_7$, we can then perform
a T-duality on the $\psi$ circle, and the corresponding dual solution
becomes
\bea
ds_{10}^2&=& e^{-\ft12 \phi}\, \Big(dx^\mu\, dx_\mu +
h^2\, dr^2 + a^2\, (\td h_1^2 + \td
h_2^2) + b^2\, (h_1^2 + h_2^2) + c^2 \td h_3^2 +
f^{-2} d\psi^2\Big)\,,\nn\\
e^{\phi}&=& f^{-1}\,,\qquad F^{\rm NS}_\3 = J_2\wedge d\psi\,.\label{nmet}
\eea
The 3-form $F^{\rm NS}_\3$ carries a magnetic 5-brane charge.  As is
typical with T-duals of regular configurations, this one is singular.
Away from the singularity, the principal orbits have the geometry of
those in the deformed conifold \cite{candel}, and share with it the
property that the $S^2$ fibres are shrinking relative to the $S^3$
base as the radius decreases.  In other words the ratio $b^2/a^2$
decreases.  Previously, a non-singular configuration using the 3-form
field strength was considered in \cite{maldnun2,bucfrey}, where the
regular 6-metric was taken to be the {\it small resolution} of the
cone over $T^{1,1}$.  This has a smoothly embedded minimal $S^2$ at
short distance (see \cite{candel,pandtsey}).  In this case it is the
$S^3$ base, rather than the $S^2$ fibres, that shrinks as the radius
decreases.  Regularity requires turning on the $SU(2)$ Yang-Mills
fields in the metric \cite{chamvolk1,maldnun2}.

   To perform the analogous $T$-duality transformation on the
fractional D2-brane, we first write the harmonic forms $G_\4$ and
$G_\3$ as
\be
G_\4=L_\4 + \wtd L_\3\wedge h_3\,,\qquad
G_\3=L_\3 + L_\2\wedge h_3\,,
\ee
where the forms $L_\4$, $\wtd L_\3$, $L_\3$ and $L_\2$ lying in the in
the 6-dimensional base manifold orthogonal to the $h_3$ fibres can be
read off directly from the expressions (\ref{4fans}) and
(\ref{f3exp}).  They are therefore given by
\bea
L_\4 &=& u_1\, a^2\, b^2\, \td h_1\wedge \td h_2\wedge h_1\wedge h_2
  + 2u_5\, h\, a\, b\, c\, dr\wedge \td h_3\wedge(\sigma_1\wedge
\sigma_2 - \Sigma_1\wedge \Sigma_2) \,,\nn\\
\wtd L_\3 &=& -u_2\, f\, a\, b\, c\, \td h_3\wedge (\td h_1\wedge
h_1 + \td h_2\wedge h_2)  + u_3\, b^2\, dr\wedge h_1\wedge h_2 +
 u_4\, a^2\,dr\wedge \td h_1\wedge \td h_2\,,\nn\\
L_\3 &=& -u_2\, h\, a\, b\, dr\wedge (\td h_1\wedge
h_1 + \td h_2\wedge h_2) + u_3\, a^2\, c\, \td h_1\wedge \td h_2\wedge
\td h_3 + u_4\, b^2\, c\, h_1\wedge h_2\wedge \td h_3\,,\nn\\
L_\2 &=& - u_1\, c\, dr\wedge (\sigma_3-\Sigma_3) + 2u_5\, f\, a\, b\,
(\sigma_1\wedge\sigma_2 -\Sigma_1\wedge \Sigma_2)\,.\label{lexp}
\eea
Then we can apply the T-duality map between type IIA
and type IIB as given in \cite{bho,clpstdual}, thereby obtaining the
following type IIB solution:
\bea
ds^2 &=& H^{-1/2}\, \Big(dx^\mu\, dx_\mu + f^{-2}\, (d\psi+m\,{\cal
B}_\1)^2\Big) + H^{1/2}\, ds_6^2\,,\nn\\
e^{\phi}&=& f^{-1}\,,\qquad F_\3^{\rm NS} = m\, (L_\3  +
L_\2 \wedge {\cal A}_\1) + J_\2\wedge d\psi\,,\label{d3brane}\\
F_\3^{\rm RR} &=& m\, \wtd L_\3\,,\qquad
F_\5 =(d^3x\wedge dH^{-1} + m\, L_\4)\wedge (d\psi + m\, {\cal B}_\1)
+ \hbox{Hodge dual}\,,\nn
\eea
where $ds_6^2$ can be read off from (\ref{eulermetric}).
Note that we have written the metric in the string frame, and that
${\cal B}_\1$ is such that $d{\cal
B}_1= L_\2$.  It is straightforward to see from (\ref{lexp}) that
${\cal B}_\1$ can be taken to be
\be
{\cal B}_\1 = -2 u_5\, f\, a\, b\, (\sigma_3-\Sigma_3)\,.
\ee
Note ${\cal B}_\1$ falls off as $1/r^2$ at large distance.  As well as
the usual electric and magnetic D3-brane charge carried by $F_\5$,
there is also magnetic 5-brane charge, since both $J_\2\wedge d\psi$
and $L_\3$ give non-vanishing flux integrals.  Like the fractional D2-brane
itself, the T-dual solution is supersymmetric, since the Killing
spinor is independent of the coordinate $\psi$ on the circle.

       It is worth mentioning that the fractional D3-brane solution
(\ref{d3brane}) can be applied more generally than just for the metric
(\ref{fourpara}) with the explicitly-constructed solution
(\ref{simsol}).  In particular, another solution of the first-order
equations obtained in \cite{bggg} for the metric (\ref{fourpara}) is
in a limit where the vector potential ${\cal A}_\1$ is effectively set
to zero.  In this limit the 1-form $h_3$ becomes simply $d\psi$, its
coefficient becomes constant, and the metric ansatz (\ref{fourpara})
reduces to an ansatz on the product of $S^1$ and six-dimensional
metrics of the Stenzel type, as discussed in \cite{cglpsten,m3brane2}.
The first-order equations for (\ref{fourpara}) then reduce to those
for the Stenzel metrics, as obtained in \cite{cglpsten}.  In this
case, which will in addition have ${\cal B}_\1=0$, the solution
becomes precisely the previously-known fractional D3-brane on the
deformed conifold \cite{klebstra}.  In our new solution, by contrast,
one direction in the world-volume space is twisted, and $F_\3^{\rm
NS}$ has an additional charge.  However the introduction of these
charges does not affect the qualitative asymptotic behaviour of $H$ at
large distance.

\section{Conclusions}

   In this paper, we have obtained the explicit solution for the
fractional D2-brane whose transverse 7-dimensional space is the new
Ricci-flat metric of $G_2$ holonomy that was constructed in
\cite{bggg}.  (This metric is asymptotically locally conical, rather
than asymptotically conical, and is analogous to one of the new ALC
metrics of Spin(7) holonomy constructed \cite{newspin7}.)  In order to
obtain the fractional D2-brane, we first constructed the appropriate
harmonic 3-form $G_\3$ in the Ricci-flat metric.  It is regular at
short distance, and the integral of $|G_\3|^2$ diverges
logarithmically in proper distance at infinity.  The equations
determining the fractional D2-brane turn out to be exactly integrable,
allowing us to obtain a fully explicit result for this solution.  It
describes D2-branes together with 5-branes wrapped over 3-cycles.

    Since the Ricci-flat $G_2$ metric obtained in \cite{bggg} is
locally asymptotic to the product of a circle of stabilised radius and
an asymptotically-conical 6-metric, one can perform a dimensional
reduction on the circle and thereby obtain a D2-brane solution in nine
dimensions with an asymptotically conical six-dimensional transverse
space.  Having first studied this reduction for the simpler examples
of vacuum solutions (Minkowski)$_4\times {\cal M}_7$ or
(Minkowski)$_3\times {\cal M}_7$, we then repeated the dimensional
reduction for the new fractional D2-brane solution.  After doing so,
by mapping into type IIB variables and then lifting back to $D=10$, we
obtained the T-dual description of the fractional D2-brane.  This has
an interpretation as a D3-brane in which one of the world-volume
directions is wrapped around the circle.

\section*{Acknowledgement}

   We are grateful to Sergei Gukov and Steve Gubser for useful
discussions and communications.  M.C.~would like to thank the
organizers of the M-theory workshop at the Institute of Theoretical
Physics at the University of California in Santa Barbara, Caltech
theory group and CERN for hospitality during the completion of this
work.  C.N.P.~is grateful to the high energy theory group at the
University of Pennsylvania, the Michigan Center for Theoretical
Physics, DAMTP, and St.~John's College Cambridge for hospitality
during different stages of this work.  Research is supported in part
by DOE grant DE-FG02-95ER40893, NSF grant No. PHY99-07949, Class of
1965 Endowed Term Chair and NATO grant 976951 (M.C.), in part by DOE
grant DE-FG02-95ER40899 (J.T.L., H.L.) and in part by DOE grant
DE-FG03-95ER40917 (C.P.).

\end{document}